\documentclass[%
reprint,
twocolumn,
superscriptaddress,
showpacs,
preprintnumbers,
graphicx,
floatfix,
byrevtex,
amsfonts,
amsmath,
amssymb,
aps,
prl,
]{revtex4-1}

\usepackage[]{graphicx}% Include figure files
\usepackage{epstopdf}
\usepackage{MnSymbol}
\usepackage{latexsym}
\usepackage{color}

\usepackage{dcolumn}% Align table columns on decimal point
\usepackage{bm}% bold math
\usepackage[mathlines]{lineno}% Enable numbering of text and display math
\begin{document}

\title{A cascading nonlinear magneto-optical effect in topological insulators}

\author{Richarj Mondal}
\affiliation{Division of Applied Physics, Faculty of Pure and Applied Sciences, University of Tsukuba, 1-1-1 Tennodai, Tsukuba 305-8573, Japan}

\author{Yuta Saito}
\affiliation{Nanoelectronics Research Institute, National Institute of Advanced Industrial Science and Technology (AIST), Tsukuba Central 5, 1-1-1 Higashi, Tsukuba 305-8565, Japan}

\author{Yuki Aihara}
\affiliation{Division of Applied Physics, Faculty of Pure and Applied Sciences, University of Tsukuba, 1-1-1 Tennodai, Tsukuba 305-8573, Japan}

\author{Paul Fons}
\affiliation{Nanoelectronics Research Institute, National Institute of Advanced Industrial Science and Technology (AIST), Tsukuba Central 5, 1-1-1 Higashi, Tsukuba 305-8565, Japan}
\author{Alexander V. Kolobov}
\affiliation{Nanoelectronics Research Institute, National Institute of Advanced Industrial Science and Technology (AIST), Tsukuba Central 5, 1-1-1 Higashi, Tsukuba 305-8565, Japan}
\author{Junji Tominaga}
\affiliation{Nanoelectronics Research Institute, National Institute of Advanced Industrial Science and Technology (AIST), Tsukuba Central 5, 1-1-1 Higashi, Tsukuba 305-8565, Japan}
\author{Shuichi Murakami}
\affiliation{Department of Physics, Tokyo Institute of Technology, 2-12-1 Ookayama, Meguro-ku, Tokyo 152-8551, Japan}

\author{Muneaki Hase}
\email{mhase@bk.tsukuba.ac.jp }
\affiliation{Division of Applied Physics, Faculty of Pure and Applied Sciences, University of Tsukuba, 1-1-1 Tennodai, Tsukuba 305-8573, Japan}

\begin{abstract}
Topological insulators (TIs) are characterized by possessing metallic (gapless) surface states and a finite band-gap state in the bulk. As the thickness of a TI layer decreases down to a few nanometer, 
hybridization between the top and bottom surfaces takes place due to quantum tunneling, consequently at a critical thickness a crossover from a 3D-TI to a 2D insulator occurs. Although such a crossover is generally accessible by scanning tunneling microscopy, or by angle-resolved photoemission spectroscopy, such measurements require clean surfaces. Here, we demonstrate that a cascading nonlinear magneto-optical effect induced via strong spin-orbit coupling can examine such crossovers. The helicity dependence of the time-resolved Kerr rotation exhibits a robust change in periodicity at a critical thickness, from which it is possible to predict the formation of a Dirac cone in a film several quintuple layers thick. This method enables prediction of a Dirac cone using a fundamental nonlinear optical effect that can be applied to a wide range of TIs and related 2D materials. 
\end{abstract}

\date{\today}

\maketitle

Higher-order nonlinear optical processes in solids are sensitive to the crystal lattice symmetry through the nonlinear susceptibilities $\chi^{(n)} (n=2, 3,...)$ \cite{Shen1}. For instance, surface second-harmonic generation in solids has been widely used as a tool to detect symmetry breaking on the surface \cite{Shen2}, which leads to non-zero $\chi^{(2)}$, while $\chi^{(2)}$ remains zero in the centrosymmetric bulk region. In addition, cascaded lower-order nonlinear processes frequently dominate over direct higher-order nonlinear processes, e.g. the cascaded second-order susceptibility leads to an effectively third order susceptibility $\chi'^{(3)} (= \chi^{(2)}\cdot\chi^{(2)})$ \cite{Kim1}, which  is in some cases greater than the direct susceptibility $\chi^{(3)}$. Thus, higher-order nonlinear optical processes on solid surfaces are complex phenomena and are often useful for exploring symmetry and have been applied to the detection of topological surface states (SSs) \cite{McIver1,Hsieh1}. 

Topological insulators (TIs) are new quantum phase of matter having  an electronic band gap in the bulk and robust SSs on the surface \cite{Zhang1,Moore,Hasan}. Bi$_{2}$Se$_{3}$ is a well-known TI while Bi$_{2}$Te$_{3}$ and Sb$_{2}$Te$_{3}$ are also considered to be model TIs although Sb$_{2}$Te$_{3}$ has been seldom investigated experimentally due to its strong $p$-type behavior stemming from point defects \cite{Hasan,Zhang2}. In the prototypical TI, Bi$_{2}$Se$_{3}$, the crossover from a 2D insulator to a 3D TI occurs at a thickness of six quintuple layers (QL), while for Bi$_{2}$Te$_{3}$ and Sb$_{2}$Te$_{3}$ this crossover is believed to occur at 3 QL and 4 QL, respectively \cite{Zhang2,Jiang,Li}. Under an external electric field, moreover, it may be possible to switch a 3D-TI to a 2D insulator or vice versa, depending on the sample thickness \cite{Kim2}. Characterization of the crossover from a 2D insulator to 3D-TI is generally difficult using optical methods where a near infrared laser pulse is applied, mainly due to the large optical penetration depth, which can result in the surface state contribution being masked by the much stronger bulk contribution. Therefore, SHG has been applied to selectively detect only the surface contribution \cite{Shi}. Here we demonstrate that nonlinear magnetic excitation and the subsequent probing of the cascaded second-order susceptibility can explicitly detect the presence of a surface Dirac cone and revisit the 2D insulator to 3D-TI crossover in X$_{2}$Te$_{3}$ (X=Bi, Sb) thin films, whose thickness are less than 10 nm. 

\begin{figure}[ht]
\centering
\includegraphics[width=\linewidth]{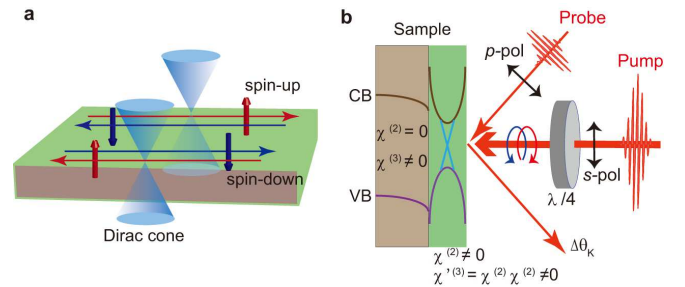}
\caption{\textbf{Schematic of the quantum spin Hall effect and the pump-probe experiment.} (\textbf{a}) In TIs, the quantum spin Hall effect occurs due to inherent strong spin-orbit coupling (SOC), which leads to edge (surface) states with two spins (spin-up and spin-down) propagating in the opposite directions. (\textbf{b}) A schematic diagram of the experiment. The TI has bulk and surface domains indicated by the shaded gray and green colors respectively. VB and CB denote the valence and conduction bands, respectively. In the bulk domain, $\chi^{(3)}\neq0$ and $\chi^{(2)}=0$, whereas $\chi^{(2)}\neq0$ appears at the surface due to the breaking of inversion symmetry at the surface. The Kerr rotational signal ($\Delta\theta_{k}$) from the TI surface states is dominated by cascading second-order nonlinear effects ($\chi'^{(3)} = \chi^{(2)}\cdot\chi^{(2)}$), which have a stronger response than a direct higher ($\chi^{(3)}$) order nonlinear process.}
\label{fig:1}
\end{figure}
Our idea is described in Fig. 1. As is well established, in TI the quantum spin Hall (QSH) effect occurs due to the presence of inherently strong spin-orbit coupling (SOC) (Fig. 1a). The QSH phase is defined as a time-reversal-symmetric system, which is gapped in the bulk while remaining gapless at the edge (surface), stemming from the system's topology \cite{Murakami}. Here we consider a TI system under irradiation by an ultrashort laser pulse (Fig. 1b).  
In general, transient magnetization $\overrightarrow{M}$ can be induced via the inverse Faraday effect (IFE) by irradiation of circularly polarized light pulses \cite{Shen1,Kimel}, which can be 
measured through the rotation of the linearly polarized light transmitted through a magnetic medium \cite{Kimel}, 
\begin{equation}
\theta_{F} = \frac{\chi}{n} \overrightarrow{M}\cdot \overrightarrow{k}, 
  \label{eqn:1}
\end{equation}
and 
\begin{equation}
\overrightarrow{M} = \frac{\chi}{16\pi}\overrightarrow{E}(\omega) \times \overrightarrow{E}^{*}(\omega), 
  \label{eqn:2}
\end{equation}
where $\chi$ is the magneto-optical susceptibility, $n$ is the refractive index of the medium, $\overrightarrow{k}$ is the wavevector, and $\overrightarrow{E}$($\omega$) is the electric field 
of the incident light with frequency $\omega$. 
The first relation, Eq. (1), corresponds to detection via the Faraday rotation and the second, Eq. (2), characterizes the pumping action, which can be regarded as optical rectification \cite{Shen1}. 
In the case of pump-probe methods, $\theta_{F} = (\chi/n)(\chi/16\pi)|\overrightarrow{E}(\omega)|^{2}$. This confirms the cascading nature of the magneto-optical susceptibility, $\chi\cdot\chi$. 
Each $\chi$ is second-order, so $\chi\cdot\chi$ is $\chi^{(2)}\cdot\chi^{(2)}$, which corresponds to $\chi^{(3)}$ (Ref. \cite{Shen1}). This means that the third-order nonlinear effect is the combination of the 
pump and probe actions\cite{Satoh}. In the case of TI materials, we can measure the transient magnetization $\overrightarrow{M}$ through the optical Kerr effect (OKE) \cite{Shen1,Hsieh2}, which is a rotation of the linearly 
polarized light reflected from the sample. In our experiment, we used an extremely thin samples of TIs. Under this ultimate condition, it is possible to detect dominant contributions from the surface, 
as explained below. In the bulk region, where $\chi^{(3)}\neq0$ and $\chi^{(2)}=0$, we observed only the usual IFE via direct OKE, while at the surface, $\chi^{(3)}\neq0$ and $\chi^{(2)}\neq0$\cite{McIver1,Giorgianni16}, leading to cascading OKE, whose amplitude can be written as $\chi'^{(3)} = \chi^{(2)}\cdot\chi^{(2)}$, can even dominate (Fig. 1b).

\section*{Results and Discussion}
Figure 2 presents the time evolution of the Kerr-rotation ($\Delta\theta_{k}$) signals for two Sb$_{2}$Te$_{3}$ samples (2 and 4 QL thick) as a function of the time delay between the pump and probe pulses under optical excitation by circularly polarized pump pulses. The $\Delta\theta_{k}$ signal changes sign upon the reversal of the helicity of the pump pulses from left to right circular polarization which are denoted by 
$\textcolor{blue}\rcirclearrowdown$  
and  
$\textcolor{red}\lcirclearrowdown$, 
respectively as shown in Fig. 2a,b. An instantaneous change in the $\Delta\theta_{k}$ signal near zero delay is observed due to the photoexcitation of spin-polarized electrons to unoccupied surface states (for 4 QL) or to bulk excited states (for 2 QL), known as the IFE \cite{Kimel,Hsieh2,Wang}. 

\begin{figure}[ht]
\centering
\includegraphics[width=85mm]{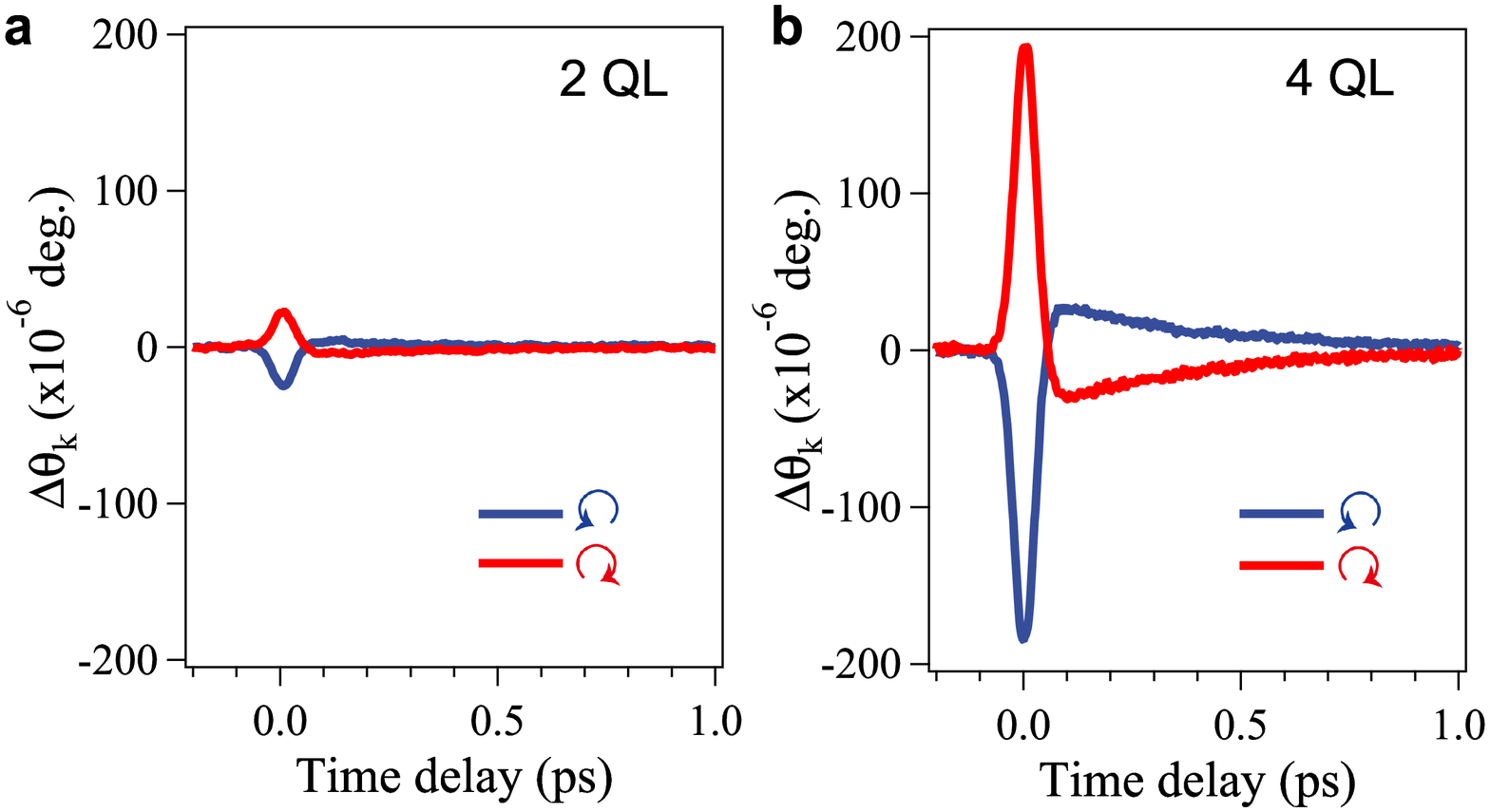}
\caption{\textbf{Time-evolution of the Kerr-rotation.} (\textbf{a}) The $\Delta\theta_{k}$ signal for 2 QL Sb$_{2}$Te$_{3}$. (\textbf{b}) The $\Delta\theta_{k}$ signal for 4 QL Sb$_{2}$Te$_{3}$. The symbol 
$\textcolor{red}\lcirclearrowdown$ 
denotes circularly right handed polarized and  
$\textcolor{blue}\rcirclearrowdown$ 
circularly left handed polarized photons.}
\label{fig:2}
\end{figure}
	
In order to investigate the critical QL thickness at which the 2D insulator to 3D-TI crossover occurs, we have measured the pump-polarization (helicity) dependence of the $\Delta\theta_{k}$ signal for Sb$_{2}$Te$_{3}$ samples with thickness in the range from 2-10 QL. Figure 3a presents the peak amplitude of the $\Delta\theta_{k}$ signal for several samples with differing numbers of QL, which exhibits a periodic oscillation with variation of the pump polarization. It is interesting to note that for the 2 QL thick sample, the oscillation displayed a two-cycle oscillation with a periodicity of $\pi$, while robust changes in the oscillatory pattern are observed for samples with thickness greater than 4 QL in the form of a four-cycle oscillation with a periodicity $\pi/2$. The abrupt change in periodicity at 4 QL clearly suggests the presence of a crossover.

\begin{figure}[ht]
\centering
\includegraphics[width=85mm]{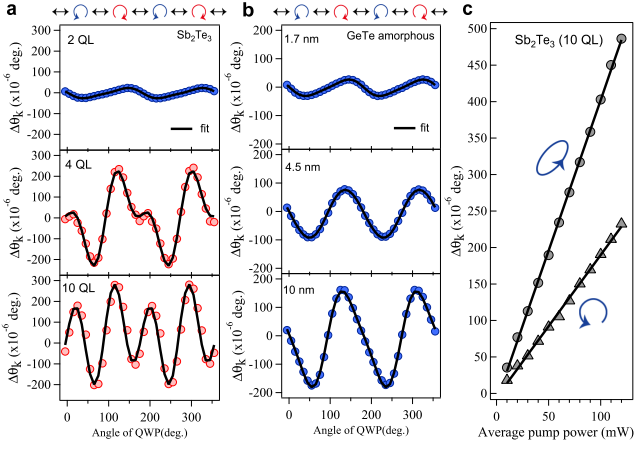}
\caption{\textbf{Helicity dependence of the Kerr-rotation signal.} (\textbf{a}) The peak amplitude of the $\Delta\theta_{k}$ signal as a function of the pump-polarization state (helicity) for Sb$_{2}$Te$_{3}$. The solid line is a fit to the data using Eq. \ref{eqn:3}. The oscillatory pattern of the Kerr rotation shows a drastic change in periodicity at 4 QL and 10 QL for Sb$_{2}$Te$_{3}$. (\textbf{b}) The same plots for amorphous GeTe films. The periodicity $(\pi)$ is the same for a 2 QL Sb$_{2}$Te$_{3}$ sample in (\textbf{a}). (\textbf{c}) The Kerr rotational amplitude measured as a function of pump power for two different QWP angles at $45^{\circ}$ and $25^{\circ}$, corresponding to circularly and elliptically polarized pump pulses, respectively. The solid line represents a linear fit. }
\label{fig:3}
\end{figure}

\begin{figure}[ht]
\centering
\includegraphics[width=65mm]{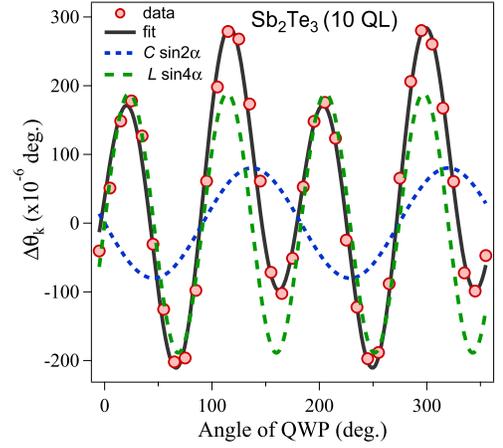}
\caption{\textbf{Helicity dependence of the Kerr-rotation signal.} The individual contribution of the specular IFE and specular OKE signals to the $\Delta\theta_{k}$ signal are shown for a 10 QL Sb$_{2}$Te$_{3}$ sample as a function of the quarter-wave plate (QWP) angle. The black solid curve is a fit to Eq. (\ref{eqn:3}). The dashed green and blue curves exhibit the contributions correspond to specular OKE and specular IFE, respectively, obtained using the fit values.}
\label{fig:4}
\end{figure}

To confirm the behavior in the periodicity of the helicity dependence of the $\Delta\theta_{k}$ signal for Sb$_{2}$Te$_{3}$ samples with a thickness of 4 QL or greater arises from a 2D insulator to 3D-TI crossover, we have carried out measurements on amorphous GeTe, a trivial insulator, for samples with a range of thicknesses (Fig. 3b). The helicity dependence of the $\Delta\theta_{k}$ signal always exhibits an oscillating pattern with a periodicity of $\pi$ with increasing amorphous GeTe thickness. The periodicity (i.e. $\pi$) of the helicity dependence of the $\Delta\theta_{k}$ signal is the same with that observed 
in crystalline GaAs\cite{Wilks}, which belongs to normal insulators as well as 
for the 2 QL Sb$_{2}$Te$_{3}$ sample (Fig. 3a). This is in accordance with the fact that topological surface states are absent in the surface of the 2 QL thick Sb$_{2}$Te$_{3}$ sample, and thus it shows similar behavior to the normal insulator GeTe.
 
The measured $\Delta\theta_{k}$ signal is linearly dependent on the pump power $I = |\overrightarrow{E}(\omega)|^{2}$ as shown in Fig. 3c. The quadratic nature of the $\Delta\theta_{k}$ signal with $|\overrightarrow{E}(\omega)|$ supports the hypothesis that the transient magnetization $\overrightarrow{M}$ induced on the TI can be treated as phenomenologically similar to a nonlinear process observed in a magnetic medium\cite{Kimel,Hsieh2,Wang}. 
Since our experimental technique utilies a reflection geometry, the observed phenomena should be given by the specular IFE and specular OKE, which have been reported to occur in semiconductor and metallic systems \cite{Popov,Svirko}. There, the semiconductor GaAs exhibited a $\pi$ oscillation, while Al metal exhibited a $\pi/2$ oscillation as a function of the polarization state (helicity) of the pump light\cite{Wilks}. The $\pi$ and $\pi/2$ periodicities were assigned to 
specular IFE and specular OKE, respectively, and the relation between $\Delta\theta_{k}$ and third-order nonlinear susceptibility $\chi^{(3)}$ was presented \cite{Popov,Svirko}. 
For the case of a TI, on the other hand, strong SOC leads to enhancement of the second-order susceptibility $\chi^{(2)}$ from the surface state \cite{Glinka15}, resulting in 
a cascading OKE via $\chi'^{(3)}={\chi^{(2)}}\cdot{\chi^{(2)}}$\cite{Kim1,Caumes,Hase}. Thus, the oscillatory pattern of the $\Delta\theta_{k}$ signal with helicity can be attributed to a combination of the specular IFE and specular OKE, or $\chi^{(3)} + \chi'^{(3)} = \chi^{(3)} + \chi^{(2)}\cdot\chi^{(2)}$. 
As such, the Kerr rotation in the case of a TI can be described by referring the expression for the specular IFE and specular OKE \cite{Popov,Svirko},
\begin{equation}
\Delta\theta_k =\frac{32\pi^2I_{pump}}{c\vert 1+n \vert^{2}} \left(L\sin4\alpha + C\sin2\alpha \right) + D,
  \label{eqn:3}
\end{equation}
where 
\begin{eqnarray}
L&=&Re\left(\frac{\chi^{(2)}\cdot\chi^{(2)}}{2n(1-n)^{2}}\right),  
 \label{eqn:4}
   \end{eqnarray}
\begin{eqnarray}   
C&=&-Im\left(\frac{\chi^{(3)}}{n(1-n)^{2}} \right).
  \label{eqn:5}
   \end{eqnarray}
Here, $\alpha$ is the angle of the quarter wave plate (QWP; $\lambda/4$ plate), $I_{pump}$ is the pump intensity, $c$ is the speed of light, and $D$ is related to a polarization-independent background. 
It should be noted that the TIs here have a rhombohedral structure with space group D$^{5}$3d (R3m)\cite{Zhang1}, leading to 81 $\chi^{(3)}$ tensor components and 27 $\chi^{(2)}$ components\cite{McIver1,Hsieh1}. Therefore, Eqs. (4) and (5) are simplified forms, rather than the reduced form available for higher symmetry crystal structures. The experimental observations can be well fit with Eq. (\ref{eqn:3}) as shown by the solid line in Fig. 3a,b. The $\pi$ period oscillation term $C\sin2\alpha$ results from the specular IFE while the $\pi/2$ oscillation component $L\sin4\alpha$ is associated with the specular OKE as the cascading OKE.

Note that the contribution of the specular OKE dominates the specular IFE in a 10 QL Sb$_{2}$Te$_{3}$ film as shown in Fig. 4. The helicity dependence of the specular IFE and specular OKE are phenomenologically similar to the circular photogalvanic effect (CPGE $\propto$ $\sin2\alpha$) and the linear photogalvanic effect (LPGE $\propto\sin4\alpha$)\cite{Ganichev} and Eq. (3) is similar to the phenomenological expression for the induced photocurrent extensively observed in TIs \cite{McIver2} and transition-metal dichalcogenides \cite{Yuan}. However, here we are observing uniquely nonlinear light-matter interactions appearing only within the duration of the optical pulse (on the order of less than 60 fs), after which an induced photocurrent will flow on a timescale of 1 ps \cite{Kastl,Braun}. 
Note that our 1.5 eV photon pump can excite electrons close to the Dirac point of the second SSs for Sb$_2$Te$_3$ \cite{Forster}, and therefore, circular-polarized pump pulse excites the spin-polarized electron population. Since our pump pulse was incident onto a sample at a slightly off-normal angle, it can create net magnetization in the second SSs. Thus, under our experimental conditions, spin-polarized photocurrents can be generated, and may be related to the CPGE and LPGE phenomena \cite{McIver1,Hsieh1}.  
Although we have provided a simple picture for the observed phenomena in TIs based on specular IFE and specular OKE, a comprehensive theoretical treatment is needed to elucidate the relationship between such nonlinear optical effects and the CPGE and LPGE phenomena on the surface of TIs. 

Similar spin dynamics have also been observed in a magnetic ionic liquid material, although the magnitude of the OKE was much smaller than that of the IFE \cite{Jin11}. We argue that the larger contribution from the specular OKE signal from the 10 QL Sb$_{2}$Te$_{3}$ film (Fig. 4) can be attributed to a surface state origin. Given that Eqs. (4) and (5) imply a half magnitude of $L$ compared to $C$, if $\chi^{(3)} = \chi'^{(3)} = \chi^{(2)}\cdot\chi^{(2)}$, our observation of $L$ with twice the magnitude of $C$ (Fig. 4) strongly indicates $\chi^{(3)} << \chi'^{(3)} = \chi^{(2)}\cdot\chi^{(2)}$, where $\chi^{(2)}$ is non-zero only in the surface region. In addition, enhancement of $\chi^{(2)}$ may be induced by inversion symmetry breaking at the surface due to Dirac plasmon-related dc electric fields \cite{Glinka15}.

\begin{figure}[ht]
\centering
\includegraphics[width=70mm]{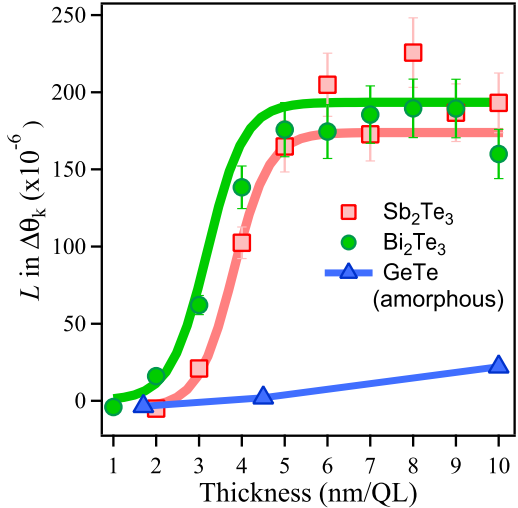}
\caption{\textbf{Detection of the crossover from a 2D insulator to a 3D-TI.} The $L$ values extracted from a fit of Eq. (\ref{eqn:3}) are plotted as a function of thickness for the TIs, Sb$_{2}$Te$_{3}$, Bi$_{2}$Te$_{3}$, and amorphous GeTe. The $L$ values for the Sb$_{2}$Te$_{3}$ and Bi$_{2}$Te$_{3}$ samples exhibit a strong variation with thickness, whereas for amorphous GeTe the signal remains constant with thickness. The thick lines represent guides for the eyes. }
\label{fig:5}
\end{figure}

\begin{figure}[ht]
\centering
\includegraphics[width=70mm]{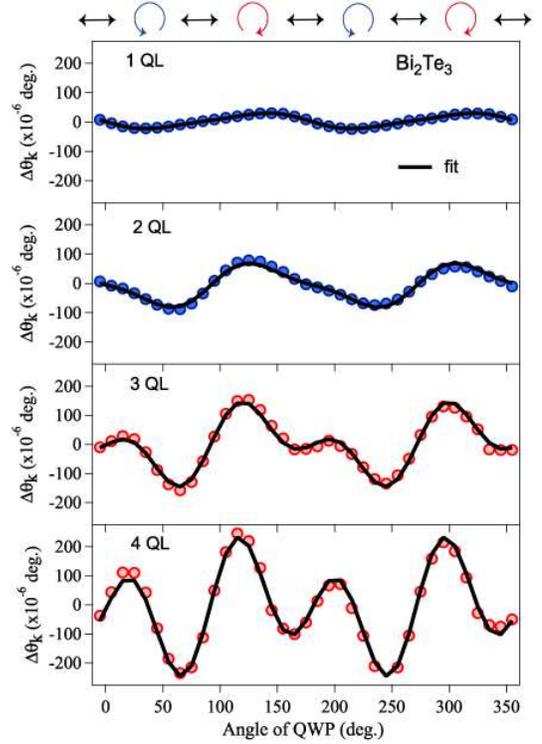}
\caption{\textbf{Helicity dependence of the Kerr-rotation signal.} The peak amplitude of the $\Delta\theta_{k}$ signal as a function of the helicity, obtained for Bi$_{2}$Te$_{3}$ samples with different thicknesses. The oscillatory pattern shows change in the periodicity from $(\pi)$ to $(\pi/2)$ at and above 3 QL for Bi$_{2}$Te$_{3}$. The solid line represents a fit to the data using Eq. (\ref{eqn:3}). }
\label{fig:6}
\end{figure}

The extracted $L$ value which can be attributed to the specular OKE is plotted as a function of thickness for Sb$_{2}$Te$_{3}$, Bi$_{2}$Te$_{3}$ and GeTe amorphous samples as shown in Fig. 5. 
Experimentally, one should thus observe an abrupt change in $L$ at 4 QL for Sb$_{2}$Te$_{3}$. This sudden change in the $L$ value indicates that a crossover from a 2D insulator to 3D-TI has occurred. We argue that the abrupt change in $L$ is associated with a change in band topology due to the presence of strong SOC, resulting in the formation of a Dirac cone \cite{Manchon}. The contribution of $L$ to the $\Delta\theta_{k}$ signal is very small for 2 and 3 QL, indicating that a gap opens at the Dirac point. Similarly, we observe an abrupt change in $L$ for 3 QL for Bi$_{2}$Te$_{3}$ (see Figs. 5 and 6). Thus, the crossover of TIs observed by the present method is consistent with the results reported in STM experiments \cite{Jiang} and \textit{ab initio} band-structure calculations (see Methods and Supplementary Fig. S1) \cite{Li}. 
On the other hand, the $L$ value almost linearly increases with sample thickness for amorphous GeTe, as it is a trivial insulator, and therefore the bulk contribution associated with $\chi^{(3)}$ increases. Therefore the contribution of the SSs ($ \chi'^{(3)} = \chi^{(2)}\cdot\chi^{(2)}$) on $L$ to the $\Delta\theta_{k}$ signal is negligible, which suggest the absence of a Dirac cone in amorphous GeTe. 
Note that the extracted $C$ value which can be attributed to the specular IFE was not correlated with the sample thickness, and therefore, we conclude that it may be due to photoinduced magnetization of the Dirac surface states of a TI, associated with $\chi^{(3)}$. 

To check if our observations are also applicable to non-topological materials, i.e., materials which exhibit SOC-induced Rashba effects\cite{Liebmann}, we have measured a crystalline GeTe sample (4.3 nm and 8.6 nm thick), where the inversion symmetry is broken due to the noncentrosymmetric lattice, i.e., $\chi^{(2)}\neq0$. As a result, we observed $\pi/2$ oscillation signals for crystalline GeTe, which is different from the case of amorphous GeTe (Fig. 3). Note that a sample at least several nm thick is required to preserve the ferroelectricity in crystalline GeTe\cite{Polking,Kolobov14}. Thus, our $\pi/2$ oscillation signals can be found in ferroelectric Rashba semiconductors, like crystalline GeTe with non-topological surface states. Therefore, specular OKE is most likely arises from strong spin-orbit coupling. 

\section*{Conclusion}
To conclude, we have experimentally explored the topological magneto-optical response from TIs using a time-resolved Kerr rotation technique. In particular,  we have focused on $p$-type Sb$_{2}$Te$_{3}$ since it is generally difficult to investigate its topological properties by conventional techniques such as ARPES. Our experimental results demonstrate that the helicity dependence of the $\Delta\theta_{k}$ signal dramatically changes when the thickness of the sample is varied. By analyzing the helicity-dependent $\Delta\theta_{k}$ signal based on the combined action of nonlinear magneto-optical effects, the specular IFE and specular OKE, we are able to discern the formation of a Dirac cone in a film several QL thick. Our method thus makes it possible the prediction of the presence of a Dirac cone (or Rashba splitting)
in air and at room temperature, i.e. without ultra-high vacuum and sophisticated equipment, using a fundamental nonlinear optical effect that can be applied to a wide range of TIs and related 2D materials. The cascading nonlinear magneto-optical effects are of particular significance for $p$-type TIs, such as interfacial phase-change materials \cite{Simpson}, which may become the foundation of the next-generation of ultra-high-speed phase change random access memory.
% (PCRAM). 

\section*{Methods}
\subsection*{Fabrication of X$_{2}$Te$_{3}$(X=Bi, Sb) films.} 
The samples used in this study were highly-oriented X$_{2}$Te$_{3}$ (X=Bi, Sb) polycrystalline films possessing a rhombohedral crystal structure with space group D$^{5}$3d (R3m)\cite{Zhang1} and a thickness between 2 to 10 QL (2 - 10 nm) for Sb$_{2}$Te$_{3}$ and 1 to 10 QL (1 - 10 nm) for Bi$_{2}$Te$_{3}$. The samples were fabricated by self-organized van der Waals epitaxy using helicon-wave RF magnetron sputtering on (100) Si substrates 
at 230$^{\circ}$C \cite{Saito}. For comparison, we prepared amorphous and crystalline GeTe films 4.3 and 8.6 nm thick, which were grown at room temperature and at 230$^{\circ}$C, respectively. 
The crystal structure of the TIs is comprised of alternating layers of X (Bi, Sb) and Te atoms along the z-direction. An alternating arrangement of five layers form one QL, which is roughly equivalent to 1 nm and has been confirmed by TEM measurements. Within each QL, X (Bi, Sb) and Te atoms are coupled by covalent bonds, whereas the coupling between QL is via van der Waals interactions. The crystal quality was checked by using coherent phonon spectroscopy\cite{Hase15}, which demonstrated the presence of both the $A_{1g}^{1}$ and $A_{1g}^{2}$ modes, which correspond to the inter- and intra-layer optical modes, respectively \cite{Richter}.
 
\subsection*{Femtosecond magneto-optical measurements.}
To investigate the critical thickness (i.e. the critical number of QL) which trigger a crossover from a 2D insulator to 3D-TI, or vice-versa, we have employed a time-resolved Kerr-rotation measurement using a novel pump-probe reflection technique. The time-resolved Kerr measurements were performed using near-infrared optical pulses with pulses generated by a Ti: sapphire laser oscillator with a pulse duration of about 20 fs, a central wavelength of 830 nm and a repetition rate 80 MHz. 
The optical penetration depth at 830 nm was estimated from the absorption coefficient to be $\sim$14 nm, which is greater than the sample thickness. Thus the optical excitation is homogeneous over the entire sample thickness, and therefore the effects of the penetration depth do not play a role on the observed crossover-like behavior in the present study. 
The average powers of the pump and the probe beams were fixed at 120 mW and 2 mW respectively. The pump and the probe beam were co-focused onto the sample to a spot size of about 70 $\mu$m with an incident angle of about $15^{\circ}$ and $10^{\circ}$ with respect to the sample normal, respectively. The maximum photo-generated carrier density induced by a single 200 $\mu$J cm$^{-2}$ pump-pulse was estimated to be $n_{exc}\approx$ 5.1$\times $10$^{19}$ cm$^{-3}$. The probe polarization was $p$-polarized while the polarization of the pump beam was modulated from linear polarized ($0^{\circ}$), to left-circular-polarized ($45^{\circ}$), to linear-polarized ($90^{\circ}$), to right-circular-polarized ($135^{\circ}$) and back to linearly-polarized ($180^{\circ}$) and so on by varying the quarter-wave-plate (QWP, $\lambda/4$ plate) angle. The change in the Kerr-rotation of the probe pulse was recorded using balanced silicon photo-diodes as a function of the pump-probe delay over a time range up to 15 ps that was introduced by a shaker operated at 19.5 Hz\cite{Hase}. The measured signal was accumulated over 700 scans to improve the signal to noise ratio. The measurements were carried out in air at room temperature.

\subsection*{Density functional theory (DFT) simulation.} 
The WIEN2k code was used for band structure calculations \cite{Schwarz}. R$_{MT}$K$_{max}$ of 7.0 was used for the plane wave component between augmentation spheres. 8$\times$8$\times$1 Monkhorst-Pack grids were used \cite{Monkhorst}. The energy convergence criterion was 0.1 mRy. The SOC term was included. The DFT-D correction was included in order to take account for the van der Waals force \cite{Grimme}.

%\bibliography{reference}
%\bibliography{sample}

\section*{Acknowledgements}
This research was financially supported by CREST (NO. JPMJCR14F1), JST, Japan. We acknowledge Ms. R. Kondou for sample preparation. 

%\section*{Author contributions}
%M. H. and J. T. planned and organized this project. Y. S. fabricated the sample. R. M. and Y. A. performed experiments and analysed the data. R. M., M. H., P. F., A. V. K., S. M. and J. T. discussed the results. R. M. and M. H. co-wrote the manuscript.

%\section*{Additional information}
%To include, in this order: \textbf{Accession codes} (where applicable); \textbf{Competing financial interests} (mandatory statement). 
%The corresponding author is responsible for submitting a \href{http://www.nature.com/srep/policies/index.html#competing}{competing financial interests statement} on behalf of all authors of %the paper. This statement must be included in the submitted article file.

%\textbf{Supplementary information} accompanies this paper at http://www.nature.com/srep\\
%\textbf{Competing interests:} The authors declare no competing interests.\\
%\textbf{Publisher's note:} Springer Nature remains neutral with regard to jurisdictional claims in published maps and institutional affiliations.

%\newpage
%\vspace{30mm}

\end{document}